\documentclass[5p]{elsarticle}
\usepackage{amssymb,amsmath}
\usepackage[english]{babel}
\usepackage{graphicx}
\usepackage{tikz}
\usetikzlibrary{patterns}

\journal{Computer Physics Communication}

\begin{document}

\begin{frontmatter}
\title{Scaling Properties of a Parallel Implementation of the Multicanonical Algorithm}
\author[]{Johannes Zierenberg\corref{cor1}}
\ead{zierenberg@itp.uni-leipzig.de}
\cortext[cor1]{Corresponding author}
\author[]{Martin Marenz}
\author[]{Wolfhard Janke}
\address{Institut f\"ur Theoretische Physik, \\
         Universit\"at Leipzig, Postfach 100920, D--04009 Leipzig, Germany}

\begin{abstract}
  The multicanonical method has been proven powerful for statistical investigations
  of lattice and off-lattice systems throughout the last two decades.
  We discuss an intuitive but very efficient parallel implementation of this algorithm and analyze
  its scaling properties for discrete energy systems, namely the Ising model and the $8$-state Potts model.
  The parallelization relies on independent equilibrium simulations in each iteration with
  identical weights, merging their statistics in order to obtain estimates for the successive weights.
  With good care, this allows faster investigations of large systems, because it distributes
  the time-consuming weight-iteration procedure and allows parallel production runs.
  We show that the parallel implementation scales very well for the simple Ising model, while the
  performance of the $8$-state Potts model, which exhibits a first-order phase transition, is limited due to
  emerging barriers and the resulting large integrated autocorrelation times.
  The quality of estimates in parallel production runs remains of the same order at same statistical cost.
\end{abstract}


\end{frontmatter}

\section{Introduction}
\label{sec:introduction}
Monte Carlo simulations are an important tool to investigate  a wide range of theoretical models
with respect to their statistical properties such as phase transitions, structure formation and more.
Throughout the last two decades, umbrella sampling algorithms like the multicanonical~\cite{BergMUCA, Janke1998}
or the Wang-Landau~\cite{WangLandau} algorithm have been proven to be very powerful for investigations
of statistical phenomena, especially first-order phase transitions, for lattice and off-lattice models.
They have been applied to a variety of systems with rugged free-energy landscapes in physics,
chemistry and structural biology~\cite{Reviews}.

Due to the fact that computer performance increases mainly in terms of parallel processing on multi-core
architectures, a parallel implementation is of great interest, if the additional cores bring a benefit to the
required simulation time.
We present the scaling properties of a simple and straightforward parallelization of the multicanonical
method, which has been reported in a similar way in \cite{ParallelMUCA} without much detail to the performance.
This parallelization considers independent Markov chains, keeping communication to a minimum.
Thus, it can be added on top of the multicanonical algorithm without much modification
or system-dependent considerations and is also suitable for systems with simple energy calculation.
Similar to this parallelization, there have been previous reports for the Wang-Landau
algorithm~\cite{Zhan2008, Landau2012}, which needed a little more adaption to the algorithm.

\section{Multicanonical Algorithm}
\label{sec:muca}
The multicanonical method allows to sample a system over a range of canonical ensembles at
the same time.
This is possible, because the statistical weights are modified in such a way that the simulation
%spends equal amounts of time in each configuration energy of a chosen interval, resulting in
reaches each configuration energy of a chosen interval with equal probability, resulting in
a flat energy histogram.
To this end, the canonical partition function, in terms of the density of states $\Omega(E)$, is modified
in the following way:
\begin{equation}
\begin{aligned}
              Z_{\rm can}  &= \sum_{\{x_i\}} e^{-\beta E(\{x_i\})} = \sum_E \Omega(E)e^{-\beta E}\\
  \rightarrow Z_{\rm MUCA} &= \sum_{\{x_i\}} W\left( E\left(\{x_i\}\right) \right) = \sum_E \Omega(E) W(E).
\end{aligned}
\end{equation}
In order that each energy state occurs with the same probability, as requested above,
the statistical weights have to equal the inverse density of states $W(E) = \Omega^{-1}(E)$.
After an equilibrium simulation with those weights, it is possible to reweight to all canonical
ensembles with a Boltzmann energy distribution covered by the flat histogram.
This can be done for example by time-series reweighting, where in the average each measured observable
is multiplied with its desired weight and divided by the weight with which it was measured:
\begin{equation}
 \langle O \rangle_{\beta} = \frac{\langle O_i e^{-\beta E_{i}}W^{-1}\left(E_{i}\right)\rangle_{\rm MUCA}}
                                  {\langle     e^{-\beta E_{i}}W^{-1}\left(E_{i}\right)\rangle_{\rm MUCA}}.
\end{equation}

Of course, the density of states and consequently the weights that yield a flat energy
histogram are not known in advance.
Therefore the weights have to be obtained iteratively.
In the most simple way consecutive weights are obtained from the last weights and the
current energy histogram, $W^{(n+1)}(E) = W^{(n)}(E)/H^{(n)}(E)$.
More sophisticated methods exist, where the full statistics of previous iterations
is used for a stable and efficient approximation of the density of states~\cite{Janke1998}.
All our simulations use this recursive version with logarithmic weights in order
to avoid numerical problems.

\subsection*{Parallel Version}
\label{sec:pmuca}
\begin{figure}[h]
  \centering
  \includegraphics[width=9cm]{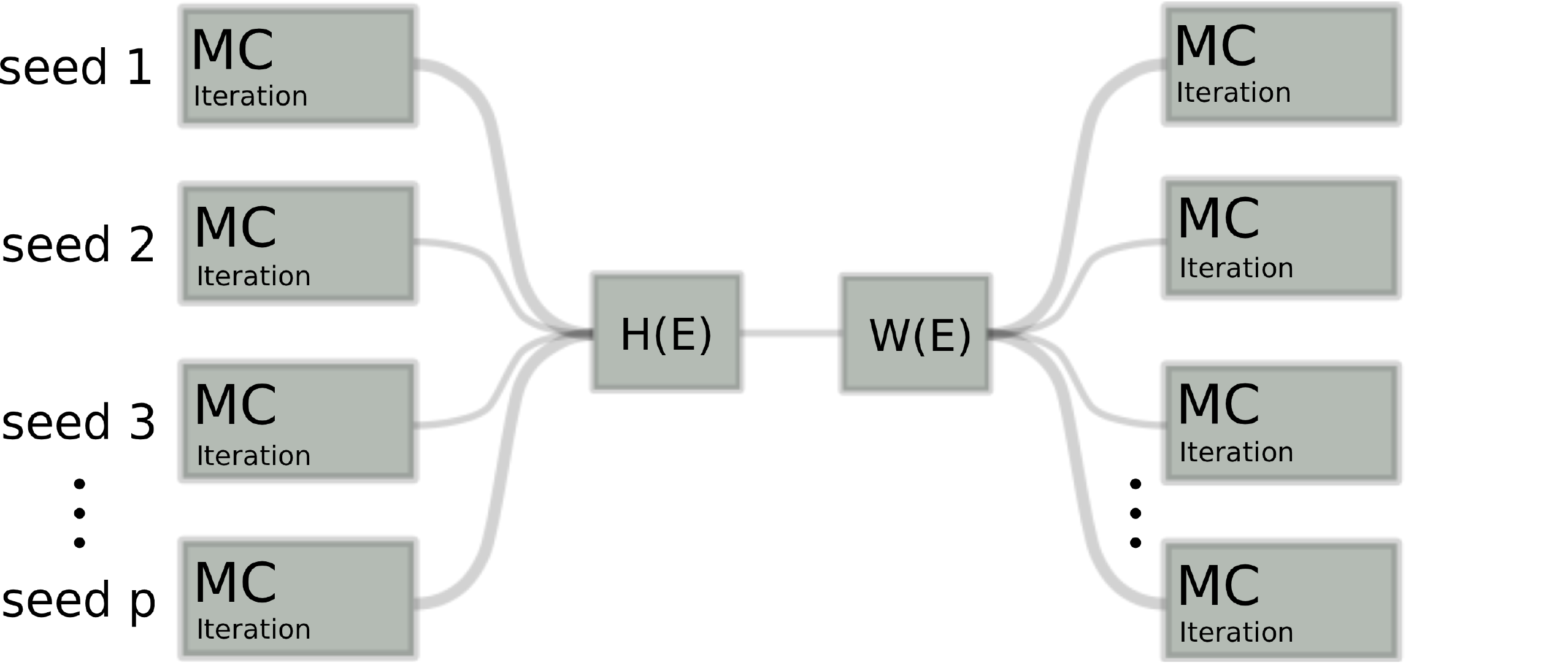}
  \caption{
            Scheme of the parallel implementation of the multicanonical algorithm on $p$ cores.
            After each iteration with independent Markov chains but identical weights, the
            histograms are merged, the new weights are estimated and the weights are distributed
            onto all processes again.
          }
  \label{fig:pmuca}
\end{figure}

The idea of this parallel implementation, similar to \cite{ParallelMUCA}, is to distribute the time
consuming generation of statistics on $p$ independent processes.
All processes perform equilibrium simulations with identical weights $W^{(n)}_{i}=W^{(n)}$, $i=1,...,p$,
but with different random number seeds, resulting in similar but independent energy histograms $H^{(n)}_{i}(E)$.
The histograms are merged after each iteration and one ends up with
 $ H^{(n)}(E) = \sum_{i}H^{(n)}_{i}(E)$.
According to the weight modification of choice, the collected histogram is processed together
with the previous weights in order to estimate the consecutive weights $W^{(n+1)}$.
The new weights are distributed onto all processes, which run equilibrium simulations again.
That way, the computational effort may be distributed on several cores, allowing to
generate the same amount of statistics in a fraction of the time.
Important to notice is that a modification of the program only influences the histogram
merging and the distribution of the new weights, see Fig.~\ref{fig:pmuca}.
The iterations are independent copies run in parallel and the weight modification is
performed on the master process as in the non-parallelized case.

\section{Systems and Implementation Issues}
\label{sec:system}
We consider two discrete two-dimensional spin systems, namely the well known Ising model
and the $q$-state Potts model with $q=8$, where the Ising model can be mapped onto the $q=2$ Potts model.
The Ising model exhibits a second-order phase transition at $\beta_0=\ln\left(1+\sqrt{2}\right)/2$ and
the $8$-state Potts model exhibits a first-order phase transition at $\beta_0=\ln\left(1+\sqrt{8}\right)$.
The spins are located on a square lattice with side length $L$ and interact only
between nearest neighbors.
In case of the Ising model, the interaction is described by the Hamiltonian
\begin{equation}
  \mathcal{H}^{\rm (Ising) } = -J\sum_{\langle i,j\rangle} s_i s_j,
\end{equation}
where $J$ is the coupling constant and $s_i$,$s_j$ can take the values
$\left\{-1, 1\right\}$.
For the $q$-state Potts model, where each site assumes values from
$\left\{0, \dots ,q-1\right\}$, the nearest-neighbors interaction is described by
\begin{equation}
  \mathcal{H}^{\rm (Potts)} = -J\sum_{\langle i,j\rangle} \delta(s_i, s_j),
\end{equation}
where $\delta(s_i,s_j)$ is the Kronecker-Delta function which is only non-zero in case $s_i=s_j$.

In those two cases the number of discrete energy states is equal to the number of lattice sites
$V=L^2$, such that the width of the energy range increases quickly with system size.
The simulations in this study start at infinite temperature, i.e., $\beta=1/T=0$ with quite narrow
energy histograms.
Because an estimation of successive weights is only possible for energies with non-zero histogram entries,
the number of iterations may be very large for wide energy ranges.
In order to ensure faster convergence, our implementation includes after each estimation of weights a
correction function, which linearly interpolates the logarithmic weights at the boundaries of the
sampled region (with a range of $L$ bins), allowing the next iteration to sample a larger energy region.
The MUCA weights are converged if the last iteration covered the full energy range and all histogram
entries are within half and twice the average histogram entry.
Between convergence of the weights and the final production run, the systems are thermalized again
in order to yield correct estimates of the observables.
In both cases, each sweep includes $V$ number of spin updates.

\section{Performance and Scaling}
\label{sec:scaling}
\begin{figure*}[t]
  \centering
  \begin{tikzpicture}
      \node at (0,0){\includegraphics{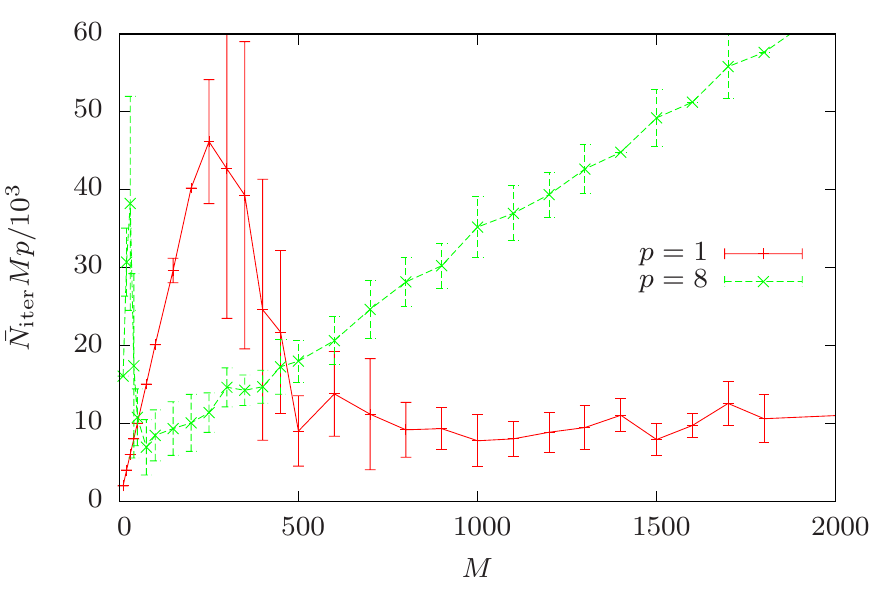}};
      \draw[rectangle, dashed, pattern=north east lines, pattern color = black!20]
            (-3.10cm,2.65cm)--(-3.10cm,-2.1cm)--(-2.80cm,-2.1cm)--(-2.80cm,2.65cm);
      \draw[rectangle, dashed, pattern=north east lines, pattern color = black!20]
            (-0.60cm,2.65cm)--(-0.60cm,-2.1cm)--(+1.30cm,-2.1cm)--(+1.30cm,2.65cm);
  \end{tikzpicture}
  \includegraphics{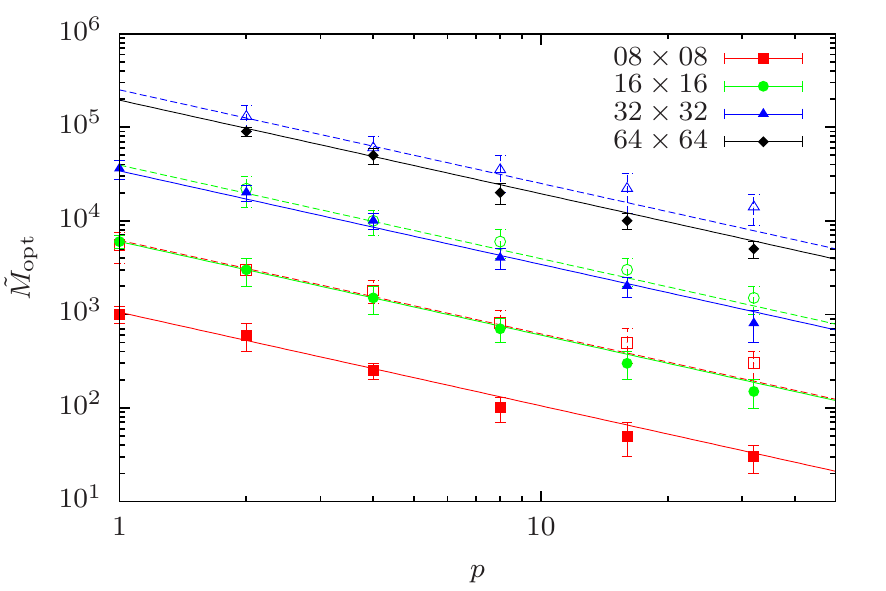}
  \caption{(left)
           Example for the estimation of $M_{\rm opt}$ for the $8\times8$ Ising model with
           $1$ and $8$ cores.
           The minimum of the total statistics $\bar{N}_{\rm iter}Mp$ with respect to $M$ yields
           $\tilde{M}_{\rm opt}$.
           (right)
           Plot of $\tilde{M}_{\rm opt}$ versus the number of cores $p$ used in the parallel MUCA simulation
           for the Ising (filled symbols) and the 8-state Potts (open symbols) model.
           The system sizes are $8\times8$ (red squares), $16\times16$ (green circles),
           $32\times32$ (blue triangles) and $64\times64$ (black diamonds).
           Fitted to the data points is the power-law dependence \eqref{eq:powerLawDependence}.
           }
  \label{fig:scalingOptimalM}
\end{figure*}

In order to estimate the performance and the speedup of the parallel algorithm appropriately,
we performed the analysis in two steps.
First, we estimated the optimal number of sweeps per iteration and core, which we will refer to
as $M$.
To this end, we performed parallel MUCA simulations over a wide range of $M$ for different
lattice sizes $L$ and number of cores $p$.
The simulations were thermalized once in the beginning on every core, continuing the next iteration
with the last state of the previous iteration on that core.
This violates the equilibrium condition a little, as no intermediate thermalization phase was applied
and part of the iteration was needed to reach equilibrium.
This is accepted in order to compare the performance equally without an additional parameter
to optimize next to $M$.
Furthermore, the physical results were not influenced, because each Markov chain was thermalized before
the final production run.
We determined the mean number of iterations until convergence to a flat histogram $\bar{N}_{\rm iter}$
as the average over $10$ simulations at different initial seeds.
Plotting the total number of sweeps $\bar{N}_{\rm iter}Mp$ versus $M$, we can estimate the optimal number
of sweeps per iteration and core $\tilde{M}_{\rm opt}$ as the minimum of this function
[see Fig.~\ref{fig:scalingOptimalM}(left)].
For a small number of cores, this curve has a rather broad minimum, introducing a rough estimate.
If, on the other hand, we stretch the curves along the $x$-axis with the number of cores, the outcomes
look quite similar.

\begin{figure}[t]
  \centering
  \includegraphics{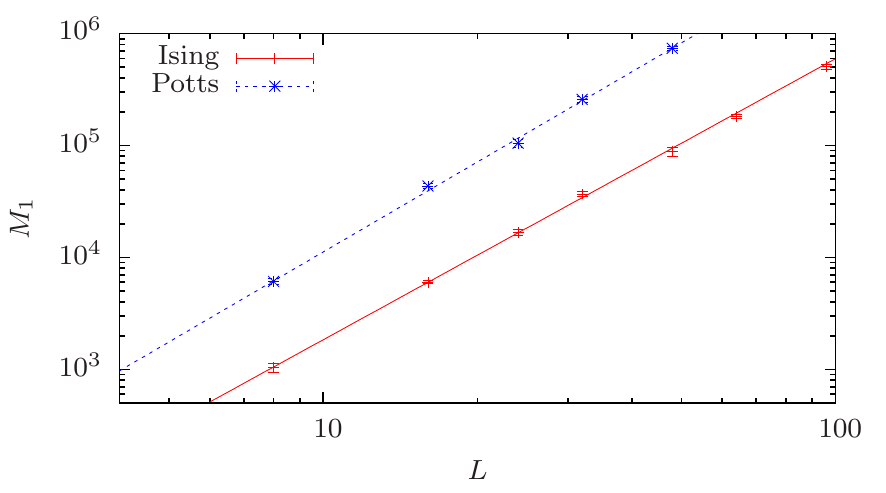}
  \caption{
           The optimal number of sweeps for a single core $M_1(L)$ together with its fit.
           Each data point is interpolated by fitting $M_{\rm opt}(p)=M_1/p$ to
           the estimated $\tilde{M}_{\rm opt}$ for $p\le32$.
           }
  \label{fig:scalingOptimalM_sizeDependence}
\end{figure}

Selected results of the estimation of $\tilde{M}_{\rm opt}$ are shown in Fig.~\ref{fig:scalingOptimalM}(right).
We see that for different lattice sizes and spin models the dependence on the number of cores
may be described by a $1/p$ power law, where the amplitudes seem to depend on the system size and
the number of states (notice that the Potts model curves nearly coincide with those of the Ising model
with $4$ times system size).
In order to measure the performance equally, it is convenient to describe $\tilde{M}_{\rm opt}$
by a function of system size $L$ and number of cores $p$,
$\tilde{M}_{\rm opt}(L,p) \approx M_{\rm opt}(L,p) = M_1(L)/p$,
where $M_1$ is the interpolated optimal $M$ for one core.
Therefore, we estimated $\tilde{M}_{\rm opt}$ for the square lattice sizes $8$, $16$, $24$, $32$, $48$, $64$, $96$
(latter two only for Ising) with $p\le32$ and fitted for fixed size $M_{\rm opt}(L,p) = M_1(L)/p$.
The obtained $M_1(L)$ were plotted over $L$ and fitted with a power law
(see Fig.~\ref{fig:scalingOptimalM_sizeDependence}).
In the end, the optimal number of sweeps per core and iteration were systematically described
by the functions
\begin{equation}\label{eq:powerLawDependence}
 \begin{aligned}
  M_{\rm opt}^{\rm (Ising) }(L,p) &= 5.7(5) \times L^{2+0.51(4)}\frac{1}{p} \\
  M_{\rm opt}^{\rm (8Potts)}(L,p) &= 24(4)  \times L^{2+0.67(6)}\frac{1}{p},
 \end{aligned}
\end{equation}
which can be justified, considering that a random walk through energy space has to depend on the
total system size and the number of spin states.
This power-law behavior corresponds roughly to the scaling of the multicanonical tunneling times
in previous works \cite{JankePotts1992, BergMUCA}.
The explicit functional dependence \eqref{eq:powerLawDependence} is characteristic for our
specific implementation and will be the basis for our performance study (including larger lattice sizes).
Interesting to notice is the prefactor ratio between $8$-state Potts and Ising (which is a $2$-Potts model)
of about $4$, corresponding to the increase in the number of spin states.

\begin{figure*}[h!t]
  \centering
  \includegraphics{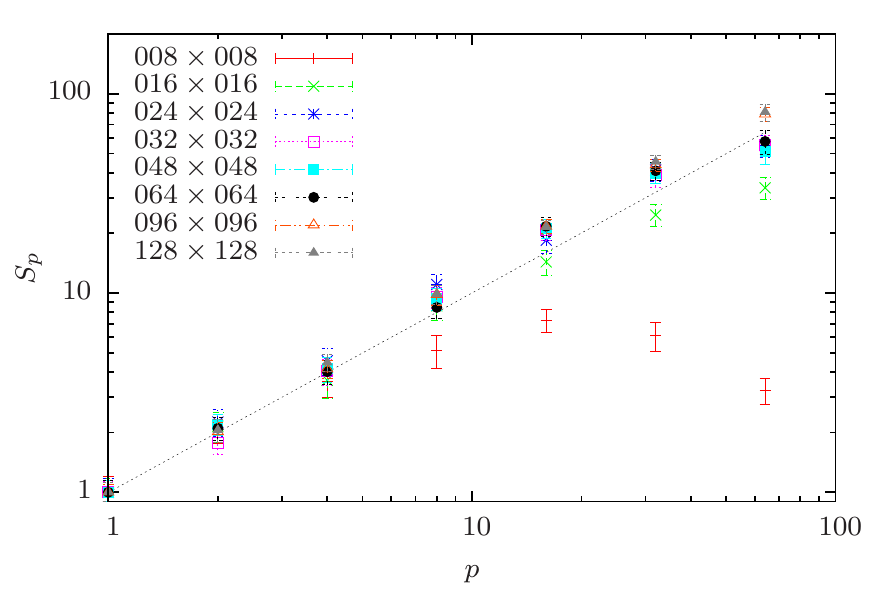}
  \includegraphics{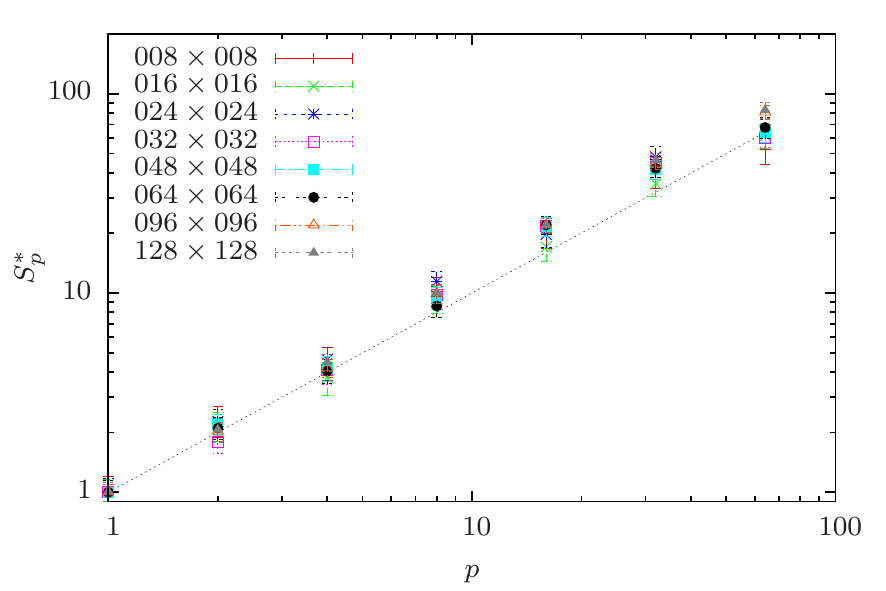}
  \caption{
           Performance for the Ising model over different system sizes expressed by
           (left)  the speedup factor and
           (right) the time-independent statistical speedup factor.
           }
  \label{fig:performance_ising}
  \vspace{1em}
\end{figure*}
\begin{figure*}[h!t]
  \centering
  \includegraphics{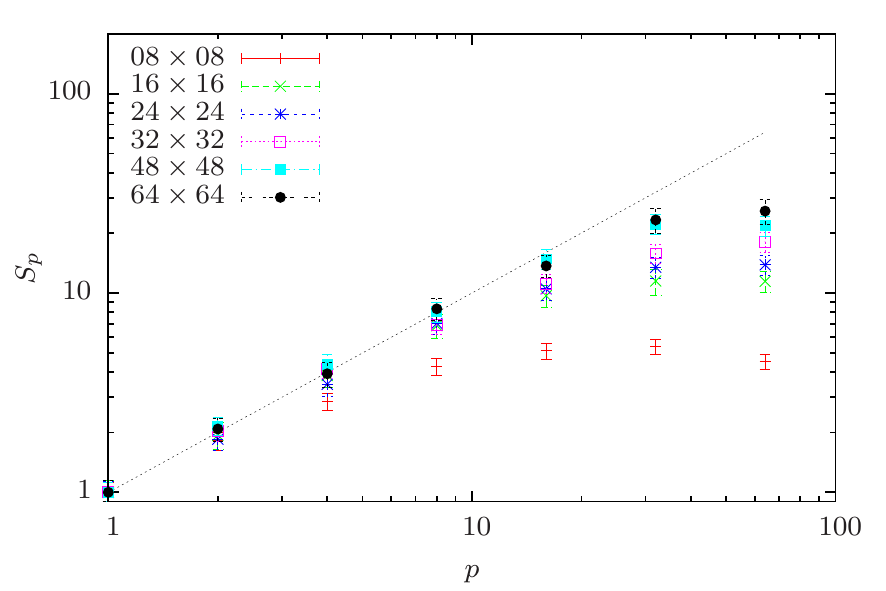}
  \includegraphics{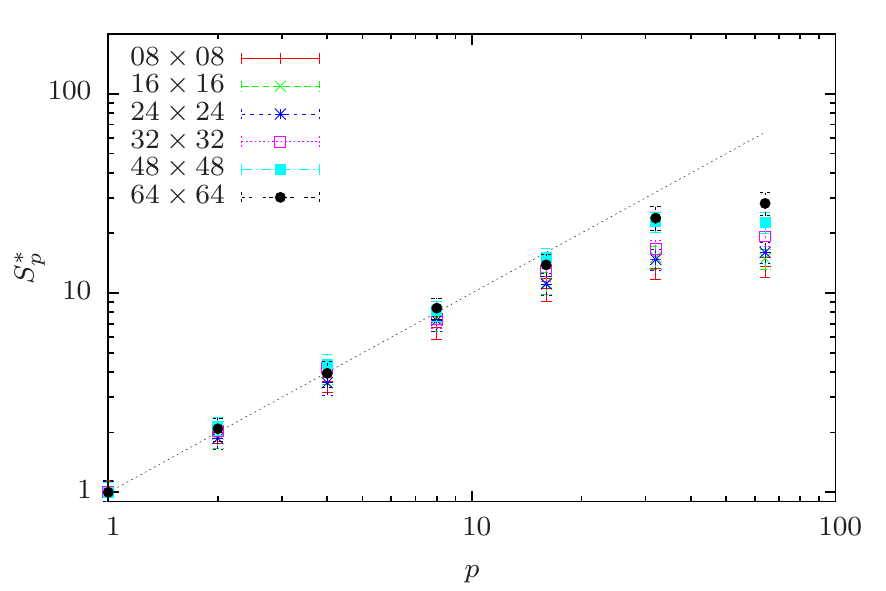}
  \caption{
           Performance for the $8$-state Potts model over different system sizes expressed by
           (left)  the speedup factor and
           (right) the time-independent statistical speedup factor.
           }
  \label{fig:performance_8potts}
  \vspace{1em}
\end{figure*}
\begin{figure*}[h!t]
  \centering
  \includegraphics{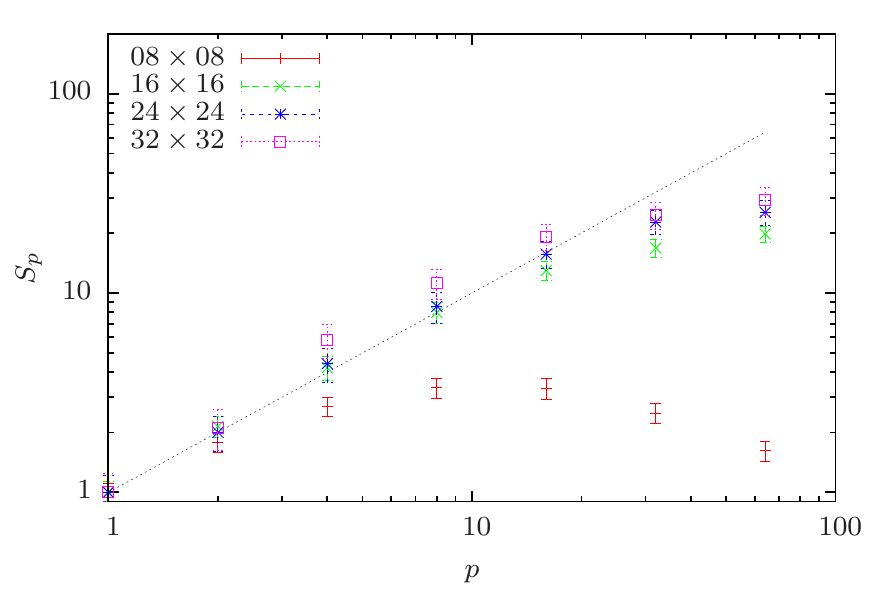}
  \includegraphics{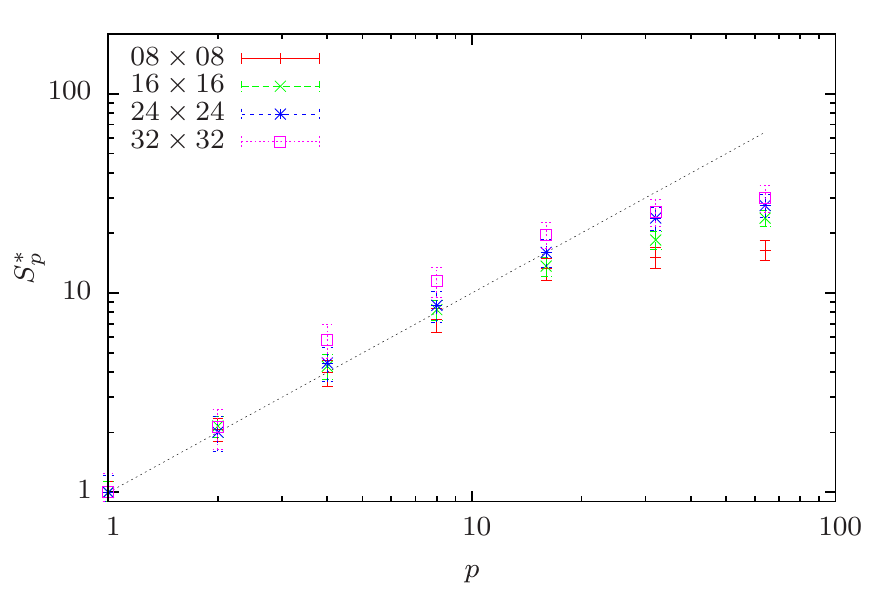}
  \caption{
           Performance for the multimagnetic simuation of the Ising model at $\beta=(3/2)\beta_0$
           over different system sizes expressed by
           (left)  the speedup factor and
           (right) the time-independent statistical speedup factor.
           }
  \label{fig:performance_isingMUMA}
  \vspace{1em}
\end{figure*}

Afterwards, we performed parallel MUCA simulations with different number of cores for each system size,
 using $M_{\rm opt}(L,p)$ in order to compare the optimal performance at each degree of parallelization.
To this end, we considered the speedup for $p$ cores, defined in terms of real time $t_p$ until
convergence of the MUCA weights,
\begin{equation}
  S_p = \frac{t_1}{t_p},
\end{equation}
as well as the time-independent statistical speedup, defined in terms of total
number of sweeps on each core until convergence $\bar{N}_{\rm iter}M_{\rm opt}(L,p)$,
\begin{equation}
    S^*_p = \frac{[ \bar{N}_{\rm iter} M_{\rm opt}(L,1)]_{1}}{[ \bar{N}_{\rm iter} M_{\rm opt}(L,p)]_{p}},
\end{equation}
where the subscript indicates the number of cores used.
In order to estimate the mean performance, we averaged the required time
and number of sweeps over $32$ independent PMUCA simulations for each data point.
The results for the Ising model are shown in Fig.~\ref{fig:performance_ising}, revealing that the
parallel implementation of MUCA results in a linear speedup up to $64$ cores already for
system sizes $L\ge24$.
In case of small system sizes $L=8$, the small $M_{\rm opt}$ leads to convergence of MUCA
weights within milliseconds, which is difficult to measure precisely.
In addition, it can be seen that the statistical speedup scales linearly for all system sizes investigated.
This means that the total required number of sweeps until convergence does not increase with the
number of cores (compare also Fig.~\ref{fig:scalingOptimalM}(left)).
This indicates no slowdown of the parallelization other than from communication, which is kept to a minimum.

In case of the $8$-state Potts model, the performance does not scale as linearly with the number of
cores, see Fig.~\ref{fig:performance_8potts}, but is still satisfying.
The reason for the drop in performance may be found in the first-order aspect~\cite{JankeFirstOrderReview} of
the Potts model.
In the transition from the disordered to an ordered phase the model undergoes a droplet-strip
transition~\cite{MartinMayor2007}.
Previous work on multimagnetic simulations of the Ising model~\cite{IsingDroplet} showed that such a transition
(and also the droplet-condensation transition) is accompanied by ``hidden'' barriers, which are not directly
reflected in the multimagnetic or multicanonical histograms and hence are difficult to overcome.
We applied the parallel version of the multicanonical method also to a multimagnetic simulation of the Ising
model at $\beta=(3/2)\beta_{0}$, determined $M_{\rm opt}(p)$ for the lattice sizes $L=8,16,24,32$ and measured
the speedup factor.
The result can be seen in Fig.~\ref{fig:performance_isingMUMA} and shows the same drop in performance
as we can observe for the $8$-state Potts model.

\begin{figure}[h]
  \centering
  \includegraphics[width=8cm]{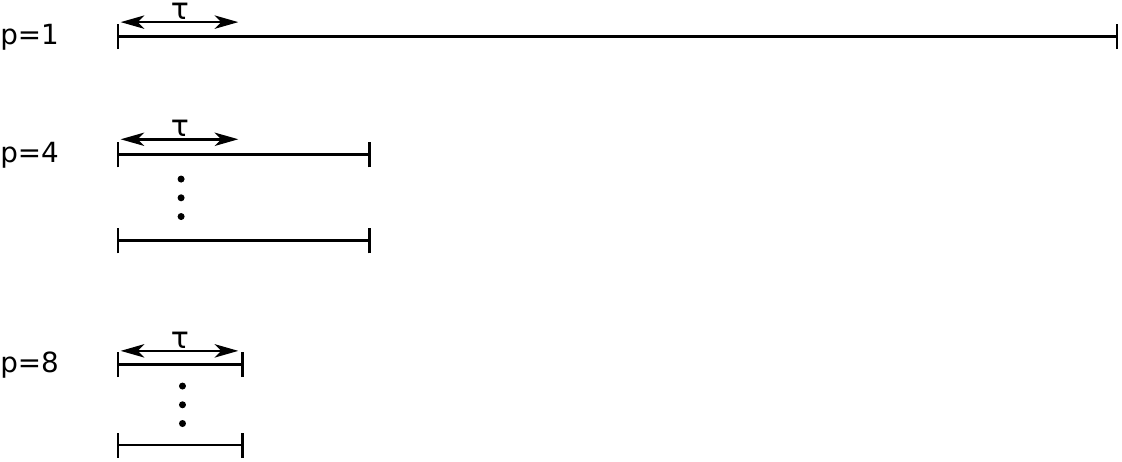}
  \caption{
            Scheme of the number of sweeps per iteration and core in comparison to the integrated
            autocorrelation time $\tau$.
            If the number of sweeps per core gets smaller than the integrated autocorrelation time
            of the Markov chain, the convergence of the MUCA weights gets worse and the performance
            drops.
            }
  \label{fig:limit}
\end{figure}

With this picture in mind, the drop in performance may be explained by the fact that with increasing $p$
we decreased the number of sweeps per iteration and thus decreased the chance to efficiently cross emerging
barriers.
When reducing the number of sweeps per iteration as a consequence of parallelization, this reaches
a point where the number of sweeps are of the order of the integrated autocorrelation time $\tau$ and each
Markov chain is strongly correlated, see also Fig.~\ref{fig:limit}.
Exemplary measurements of $\tau$ in the $8$-state Potts model for different lattice sizes revealed that
it was of the order of $M_{\rm opt}(L,8)$ to $M_{\rm opt}(L,16)$, verifying the drop in performance for $p\ge16$.
This gives a limit of parallelization depending on the barriers and the associated autocorrelation times
of the system.

From a practical point of view, when simulating complex systems, it may be advantageous to introduce
short thermalization phases between iterations.
Exemplary investigations of the $8$-state Potts model with intermediate thermalization showed that
the number of iterations can be reduced significantly while introducing additional computation time.

\section{Quality}
\label{sec:quality}
\begin{figure*}[t]
  \centering
  \includegraphics{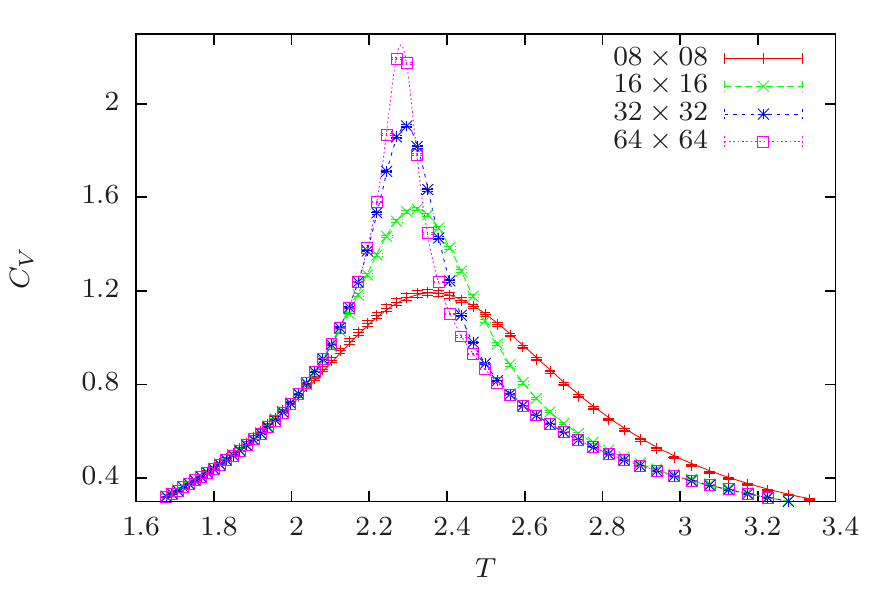}
  \includegraphics{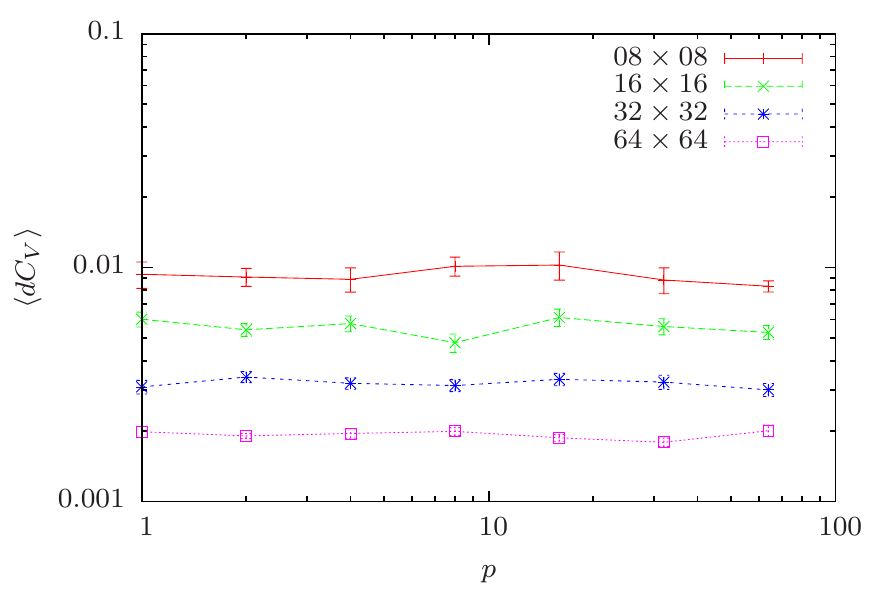}
  \caption{
           (left) Reweighted specific heat from an Ising simulation with $64$ cores and the exact
           results as well as
           (right) its relative deviation from the exact solution over the number of
           cores with constant total number of measurements.
          }
  \label{fig:quality_ising}
  \vspace{1em}
\end{figure*}

%%%%%Ising
The parallel MUCA weight recursion can be extended by a parallel production run, acquiring data
with independent Markov chains.
This allows equally good estimation of observables for a constant number of total measurements,
if it is ensured that each Markov chain samples the desired phase space appropriately.
For the Ising model, we considered the relative deviation from the exact result
$O_{\rm ex}$~\cite{Kaufman1949} and averaged over a temperature range around the critical temperature
\begin{equation}
  \langle dO \rangle = \left(1/N\right) \sum_{\beta_i} \frac{|O(\beta_i)-O_{\rm ex}(\beta_i)|}{O_{\rm ex}(\beta_i)}.
\end{equation}
In this case the range was chosen to be $\beta\in[3/10,6/10]$ with $N=300$ steps.
Figure~\ref{fig:quality_ising} shows the average deviation of the specific heat
$\langle dC_V \rangle$ ($C_V=\beta^2\left(\langle E^2\rangle - \langle E\rangle^2\right)/V$) for
different sizes of the Ising model.
The error was estimated by averaging over $16$ independent runs.
In each simulation there was an additional thermalization phase after the MUCA-weight convergence
and the final production run was performed with $30\times M_{\rm opt}(L,p)$ measurement points
and $50$ sweeps between measurements.
The decrease of the relative deviation with increasing lattice size comes from this choice.
From Fig.~\ref{fig:quality_ising}(right) it can be seen that, for a given system size,
the relative deviations remain constant within the statistical error for all $p$.
%

\begin{figure*}[t]
  \centering
  \includegraphics{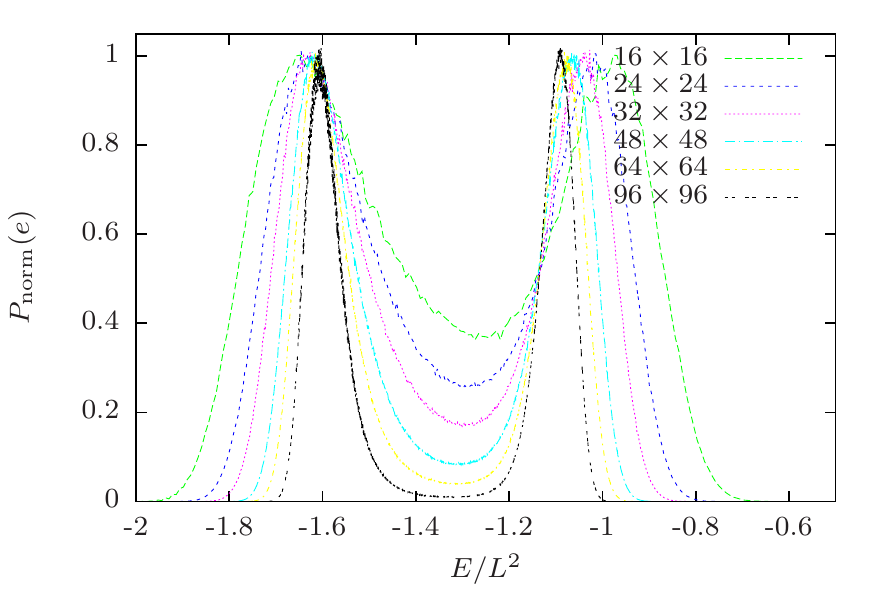}
  \includegraphics{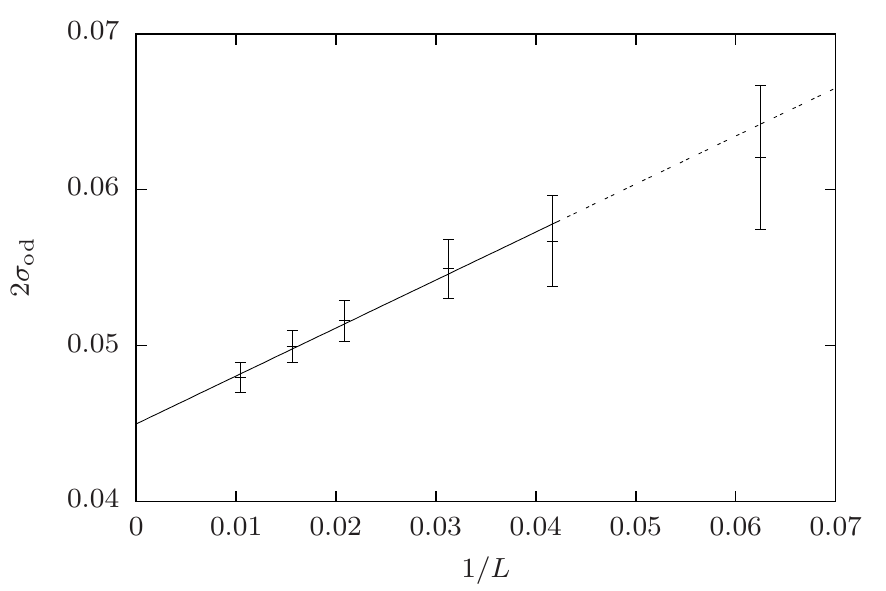}
  \caption{
            (left)  Reweighted normalized mean energy histograms and
            (right) the scaling of the order-disorder interface tension for the
            $8$-state Potts model simulated on $64$ cores.
          }
  \label{fig:quality_potts}
  \vspace{1em}
\end{figure*}

%%%%%Potts
In order to verify the quality of the parallel simulation of the $8$-state Potts model, we estimated
the scaling of the order-disorder interface tension $\sigma_{\rm od}$ and compared it to analytic
results~\cite{JankeInterfaceTension1992}.
The interface tension can be approximated in terms of the probability density of the histograms at the
transition temperature~\cite{InterfaceTension},
\begin{equation}
  \sigma_{\rm od} = \lim_{L\rightarrow\infty} \frac{1}{2 L}\ln\left(\frac{P_{\rm max}}{P_{\rm min}}\right)_{\beta_{0}}.
\end{equation}
This requires to first find the temperature for which the reweighted energy histogram
shows two equally high peaks [see Fig.~\ref{fig:quality_potts}(left)] and then to estimate the ratio
between the maximum and minimum.
Figure~\ref{fig:quality_potts}(right) shows a rough scaling of the order-disorder interface tension for
several system sizes up to $96\times96$, simulated with $64$ cores.
The fit to the larger system sizes yields an infinite lattice limit \mbox{$\sigma_{\rm od}\approx0.045$},
which is consistent with the exact result~\cite{JankeInterfaceTension1992} and verifies that
the parallel implementation yields correct results.

\section{Conclusion}
\label{sec:conclusion}

We presented a straightforward parallel implementation of the multicanonical algorithm and
showed that its scaling properties with the number of cores are very good for the Ising model
and adequate for the $8$-state Potts model.
The latter one suffers from emerging barriers at the first-order phase transition,
resulting in large integrated autocorrelation times.
The parallelization profits from a minimal amount of communication because histograms are
merged at the end of each iteration.
This is the main reason why it is difficult to adapt this method to weight recursions of the
Wang-Landau type where the weights are changed after each spin update.
Since it would be interesting to find a similar approach for those weight recursions,
not relying on shared memory, we are currently investigating this problem.
The major advantage of the implementation employed here lies in the fact, that no greater adjustment
to the usual implementation is necessary and that additional modifications may be carried along.
Thus, it can be easily generalized to complex systems, e.g. (spin) glasses or (bio) polymers,
and allows good convergence if it is ensured that the number of sweeps per core is greater than
the integrated autocorrelation times.

\section*{Acknowledgments}
We are grateful for support from
 the Leipzig Graduate School of Excellence GSC185 ``BuildMoNa'' and
 the Deutsch-Franz\"osische Hochschule (DFH-UFA) under grant No.\ CDFA-02-07.
This project was funded by the European Union and the Free State of Saxony.

\section*{References}
\bibliographystyle{model1-num-names}

\end{document}